\documentclass[prl,twocolumn,amsmath,amssymb,superscriptaddress,showpacs]{revtex4-1}

\usepackage{natbib}
\usepackage{graphicx}
\usepackage[usenames]{color}

\newcommand{\fig}[1]{Fig.~\ref{#1}}
\newcommand{\be}[1]{\begin{equation}\label{#1}}
\newcommand{\ee}{\end{equation}}

\begin{document}
\title{Attosecond e$^{-}$- e$^{-}$ collision dynamics of  the four-electron escape  in Be close to threshold}
\author{A. Emmanouilidou$^{1,2}$ and H. Price$^{ }$}

\affiliation{$^{ }$Department of Physics and Astronomy, University College London, Gower Street, London WC1E 6BT, United Kingdom\\
$^2$Chemistry Department, University of Massachusetts at Amherst, Amherst, Massachusetts, 01003, U.S.A.}

\begin{abstract}
We explore the escape geometry of four electrons a few eV above threshold following  single-photon absorption from the ground state of Be. We find that the four electrons leave the atom on the vertices of a {\it pyramid} instead of a previously-predicted tetrahedron. 
 To illustrate the physical mechanisms of  quadruple ionization we use a {\it momentum transferring attosecond collision} scheme which we show to be in accord with the {\it pyramid} break-up pattern. 
\end{abstract}
\pacs{32.80.Fb, 34.80.Dp, 05.45.-a, 41.50.+h}
\maketitle

Exploring the correlated electronic motion during ionization of multi-electron atoms and molecules, for energies close to the ionization threshold, is a fundamental and challenging task in physics. 
This electronic correlation has been a topic of intense interest, for most recent see ref. \cite{Exp1,Exp2}, since Wannier's pioneering work \cite{wan53}. According to Wannier's law
   $\mathrm{\sigma\propto E^{\beta}}$ for excess energies $\mathrm{E\rightarrow 0}$, where $\mathrm{\sigma}$ is the cross-section of the process involved and $\mathrm{\beta}$ 
   depends on the number and type of particles involved in the break-up process. Using classical mechanics, 
Wannier also
showed that  two electrons moving in the Coulomb field of an ion  escape back-to-back for energies $\mathrm{E\rightarrow 0}$. Extending Wannier's work, later studies  
predicted a three-electron break-up on the vertices of an equilateral triangle  \cite{ks76,Ostrovsky1998} and a four-electron break-up on the vertices of a  tetrahedron  \cite{Ostrovsky1998,Crujic}. While these highest-symmetry break-up
patterns were predicted for $\mathrm{E\rightarrow 0}$ it is  generally expected that they also prevail for excess energies  a few eV above threshold where the threshold Wannier exponent $\mathrm{\beta}$
is still retrieved. In this work we show  that this  is not true.

We show that for single-photon quadruple ionization (QI) from the ground state of Be the prevailing break-up pattern a few eV above threshold is different than the one predicted for $\mathrm{E\rightarrow 0}$.
For single-photon triple ionization  from the ground state of Li we have already shown that the break-up pattern is not the expected ``triangle" but a {\bf T}-shape  a few eV above threshold     \cite{Emmanouilidou2006JPB}. In the {\bf T}-shape two electrons escape back-to-back while the third electron escapes  
 at $90^{\circ}$ with respect to the other two electrons.
Very recently, further evidence for the ${\bf T}$-shape was provided by fully quantum mechanical calculations for energies 5 eV above the triple ionization threshold of the ground state of Li  \cite{ColganA}.
 The previously predicted  ``triangle" pattern was, however, observed in recent $(e,3e)$ coincidence measurements for  
electron-impact on the ground state of He \cite{ren08}.
 The above reinforce a prediction we made in \cite{ewr08} that the three-electron break-up pattern depends on the initial state 
and can be either a ${\bf T}$-shape or a ``triangle".  

In the current work, we present evidence that for single-photon QI from the ground state of Be, a few eV above threshold, the prevailing break-up pattern is a {\it pyramid}. That is, the three electrons  escape on the vertices of an equilateral triangle at 120$^{\circ}$ from each other and the other electron escapes perpendicular to the plane of the three electrons. Our prediction differs from the symmetric four-electron escape on the vertices of a  tetrahedron  predicted in the limit $\mathrm{E\rightarrow 0}$ \cite{Ostrovsky1998,Crujic}. Moreover, uncovering the physical mechanisms of QI, we express the multi-electron escape dynamics in terms of {\it momentum transferring attosecond collision sequences}. Thus, besides addressing a fundamental law of physics, we also elucidate correlated electronic motion in multi-electron escape. This is of high interest since the advent of ultrashort and
intense laser pulses has brought time-resolving correlated electron dynamics in intra-atomic ionization processes  at the forefront of Attosecond Science   \cite{KlunderPRL2011, Taylor, Uiberacker2007Nature, Eckle2008Science,Schultze2010Science}.

 Given computational capabilities, addressing four-electron escape with quantum mechanical techniques is currently out of reach \cite{Colgan}. Classical mechanics is justified for excess energies close to threshold as detailed in the original work of Wannier \cite{wan53} and in subsequent work on two \cite{RostTobias} and three-electron escape \cite{Emmanouilidou2006JPB, ewr08}. We tackle quadruple photoionization  using the quasiclassical technique---quasi due to the choice of initial state---detailed in \cite{EmmanouilidouBe}. Briefly, using the Classical Trajectory Monte Carlo method \cite{CTMC1,CTMC2}, we propagate in time the full five-body Hamiltonian accounting for all interactions to all orders. In addition, we use a Wigner \cite{Wignertran} distribution for setting-up the initial phase space of the bound electrons.  We compute the probability for QI, $\mathrm{P^{4+}}$, for excess energies ranging from  3  eV to 10 eV. 3 eV  is  close to threshold and the computational time  involved is not prohibitive for obtaining good statistics. 10 eV  is an  upper bound estimate of excess  energies where the Wannier exponent $\mathrm{\beta}$  can still be retrieved. Using our data for $\mathrm{P^{4+}}$ from 3 eV to 10 eV in steps of 1 eV we find $\mathrm{\beta}$ equal to 94\% of the theoretically predicted value of 3.288 \cite{Ostrovsky1998}. In the framework of Wannier's theory, in what follows   we discuss our results for 3 eV and 10 eV.

\begin{figure}
  \includegraphics[width=0.35\textwidth]{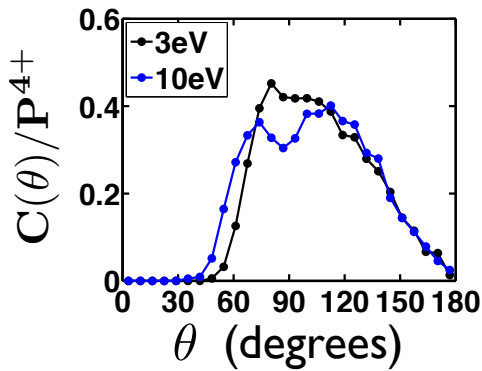}           
 \caption{Probability for two electrons to escape with an inter-electronic angle $\theta$ for excess energies 3 eV (black dots) and 10 eV (blue dots). To guide the eye, for each excess energy, we connect the dots representing our data with a line.}
  \label{fig:theta}
\end{figure}

To identify the four-electron escape pattern we focus on an observable that naturally encompasses electronic correlation. Such an observable is the probability for two electrons to escape with an inter-electronic angle $\mathrm{\theta}$---we refer to it as angular correlation probability $\mathrm{C(\theta)}$. 
In \fig{fig:theta}, we plot $\mathrm{C(\theta)}$ for 3 eV and 10 eV excess energies. Given that $\mathrm{P^{4+}}$ is 1.8$\times$10$^{-10}$ for 3 eV and 7.3$\times$10$^{-9}$ for 10 eV the computational task involved is immense. Nevertheless, to provide good accuracy, for each excess energy we consider, our results involve roughly 10$^4$ quadruple ionization events. In \fig{fig:theta}, we see that  for 10 eV  $\mathrm{C(\theta)}$ has two peaks: one around 74$^{\circ}$ and a second one around 100$^{\circ}$-125$^{\circ}$. However, for 3 eV it is not clear whether  only one or two less---compared to 10 eV---pronounced peaks are  present in the range 80$^\circ$-112$^\circ$.  In \fig{fig:theta}, $\mathrm{C(\theta)}$  is plotted using 28 bins for $\mathrm{\theta}$. We choose the bin size so that the double peak structure in $\mathrm{C(\theta)}$ is best resolved given the limitations imposed by our statistics. 

To what four-electron escape geometry does the shape of $\mathrm{C(\theta)}$  correspond to?  A tetrahedron pattern with all electrons escaping at 109.5$^\circ$ from each other would result in a single peak in $\mathrm{C(\theta)}$. A {\it pyramid} pattern  with three electrons escaping at 120$^\circ$ from each other and the other electron escaping  at 90$^\circ$ from the three electrons would result in two peaks in $\mathrm{C(\theta)}$. 
Hence, the double peak in $\mathrm{C(\theta)}$ (\fig{fig:theta}) for 10 eV is more consistent with a {\it pyramid}-shape while  for 3 eV the shape of  $\mathrm{C(\theta)}$ does not provide conclusive evidence for the prevailing escape geometry.

 We next elucidate the physical mechanisms of QI and provide conclusive evidence for the  break-up patterns the four electrons follow. How does the photo-electron redistribute the energy it gains from the photon to the remaining three electrons? This is a natural question in the framework of classical mechanics where the electrons undergo soft collisions mediated by Coulomb forces. Does redistribution of energy take place through one simultaneous collision between all 4 electrons or through a sequence of collisions? To answer this question, we use a classification scheme similar to the one we first introduced in the context of three-electron escape following single-photon absorption from the ground state of
 Li \cite{Emmanouilidou2006JPB}. That is, we define a collision between electrons i and j---labeling it as $\widehat{ij}$---through the momentum transfer 
 \begin{equation}
  \mathrm{{\bf D}_{ij}=\int_{t_{1}}^{t_{2}} {\bf \nabla} V(r_{ij}) dt} 
  \end{equation}
 under the condition that $\mathrm{V(r_{ij}(t_{k}))}$ are local minima in time with $\mathrm{t_{2}>t_{1}}$ while $\mathrm{r_{ij}=|\bf{r}_{i}-\bf{r}_{j}|}$ and $\mathrm{V(r_{ij})=1/|\bf{r}_{i}-\bf{r}_{j}|}$. During the time interval $\mathrm{t_1<t<t_{2}}$ all five particles interact with each other. Hence, the above definition is meaningful if the collision redistributes energy primarily within the three-body subsystem that includes the nucleus and the electrons i and j. For automated identification of the collisions, we need sensitivity thresholds to register only the important collisions for the quadruple events. We do so for each individual QI trajectory by forming the maximum  
$\mathrm{D=max_{i\neq j}|{\bf D}_{ij}|}$ 
and registering only collisions with $\mathrm{|{\bf D}_{ij}|/D}>\delta$ where $\mathrm{i,j=1,2,3,4}$. 
We introduce another sensitivity  threshold for how ``sharp" a collision is. Namely, if electron i gains energy through more than one  collisions, we find the maximum  $\mathrm{\Delta V_{i}=max_{i\neq j}(V(r_{ij})^{max}-V(r_{ij})^{min})}$, with $\mathrm{V(r_{ij})^{max/min}}$ the max/min value of  $\mathrm{V(r_{ij}(t))}$ for $\mathrm{t_{1}<t<t_{2}}$, and register only collisions satisfying $\mathrm{(V(r_{ij})^{max}-V(r_{ij})^{min})/\Delta V_{i}>\delta_{1}}$. We have checked that our results and conclusions do not change for different  values of  $\mathrm{\delta}$ and $\mathrm{\delta_{1}}$; we choose $\mathrm{\delta=1/12}$ and $\mathrm{\delta_{1}=1/8}$.

 \begin{figure*}
\includegraphics[width=1\textwidth]{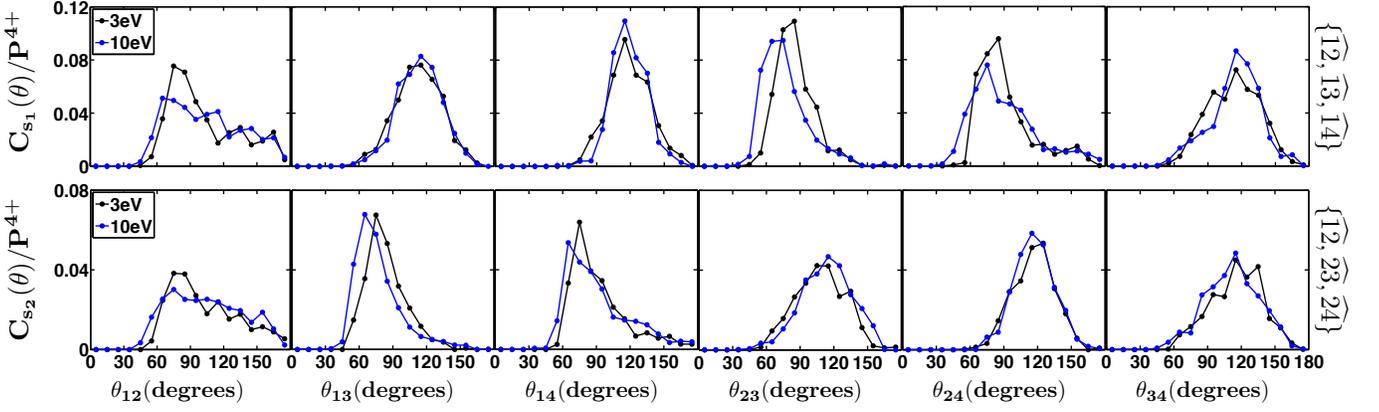}
\caption{Same as \fig{fig:theta} but  for each inter-electronic pair  $\mathrm{\theta_{ij}}$ for the ionization routes $\mathrm{s_{1}}$ (top row) and $\mathrm{s_{2}}$ (bottom row).  }
\label{fig:collisions}
\end{figure*}

According to this classification scheme we find that electrons 2, 3 and 4 gain sufficient energy to leave the atom through two prevailing ionization routes. We denote by electron 1 the photo-electron (from an 1s orbital), by 2 the other 1s  electron and by 3 and 4 the two 2s electrons.
 In the first route the photo-electron 1 knocks-out 
first electron 2 and then proceeds to knock-out electrons 3 and 4. That is, first a collision $\widehat{12}$ takes place very early in time and roughly 24 attoseconds later collisions $\widehat{13}$ and $\widehat{14}$ occur. With collisions $\widehat{13}$ and $\widehat{14}$ taking place close in time we find that a fourth collision $\widehat{34}$ can occur in addition to the previous 3 collisions.  We refer to this ionization route where the photo-electron transfers energy to both electrons 3 and 4 as  $\mathrm{s_{1}=\{\widehat{12},\widehat{13},\widehat{14}\}}$.   In the second route, the photo-electron 1 first knocks-out electron 2 through the collision $\widehat{12}$. Then, electron 2 becomes the new impacting electron knocking-out, roughly 24 attoseconds later, electrons 3 and 4 through the collisions $\widehat{23}$ and $\widehat{24}$.   With collisions $\widehat{23}$ and $\widehat{24}$ taking place close in time a fourth collision $\widehat{34}$ can occur in addition to the previous 3 collisions.   We refer to this ionization route where electron 2 transfers energy to both electrons 3 and 4 as  $\mathrm{s_{2}=\{\widehat{12},\widehat{23},\widehat{24}\}}$.  $\mathrm{s_{1}}$ accounts for 41\% and  $\mathrm{s_{2}}$ for 24/26\% of all QI events for 3/10 eV. Using this scheme of {\it momentum transferring attosecond collision} sequences we have thus obtained a physical picture of the correlated electronic motion in an  intra-atomic ionization process. Further, this scheme offers insight in choosing the appropriate asymptotic observables  for inferring the temporal profile of electron-electron collision dynamics \cite{Agapistreak}. This is important since developing pump-probe schemes to time-resolve correlated multi-electron escape is one of the current challenges facing Attoscience \cite{ReviewAtto}.

\begin{figure}[h]
  \includegraphics[width=0.5\textwidth]{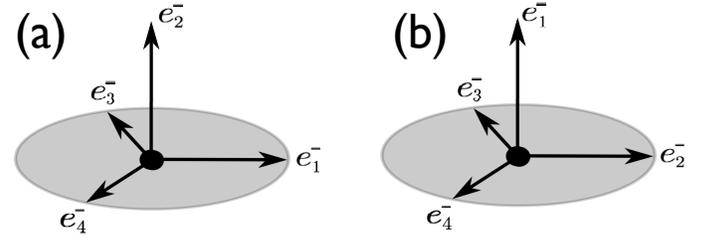}           
 \caption{The {\it pyramid} escape geometry for four electrons corresponding to collision sequences $\mathrm{s_{1}}$ (a) and  $\mathrm{s_{2}}$ (b).}
 \label{fig:break-up}
\end{figure}

 In what direction do the four electrons escape in pathways $\mathrm{s_{1}}$ and $\mathrm{s_{2}}$?  For $\mathrm{s_{1}}$, at the time when all electrons to be ionized have received enough energy to leave the atom the spatial electron distribution, we refer to it as transient threshold configuration TTC \cite{ewr08}, is  $\mathrm{r_{1}\approx r_{3} \approx r_{4} \neq r_{2}}$. That is, the last colliding electrons 1, 3 and 4 have $\mathrm{r_{1}\approx r_{3} \approx r_{4}}$ which is 
 close to the fixed point (see below) of the four-body Coulomb problem---three electrons and the nucleus. Thus, one  expects that  electrons 1, 3 and 4 will escape symmetrically on a plane at 120$^{\circ}$ from each other.
  In \fig{fig:collisions} (top row) we plot  $\mathrm{C(\theta)}$ for each of the six inter-electronic angles of escape using only the QI events corresponding to the $\mathrm{s_{1}}$, i.e, we plot $\mathrm{C_{s_{1}}(\theta)}$. Indeed, we see that $\mathrm{C_{s_{1}}(\theta)}$ for $\mathrm{\theta_{13}}$,  $\mathrm{\theta_{14}}$ and $\mathrm{\theta_{34}}$ peak around 115$^{\circ}$, both for 3 eV and 10 eV, corresponding to electrons 1, 3 and 4 escaping on the vertices of a ``triangle". (We note that the distributions in \fig{fig:theta} and \fig{fig:collisions} are convoluted by  the polar angle volume element $\mathrm{\sin{\theta}}$ resulting in a peak at 120$^{\circ}$ being shifted to slightly smaller angles while a peak at 90$^{\circ}$ is not affected). In addition, we see that  $\mathrm{C_{s_{1}}(\theta)}$ for
$\mathrm{\theta_{12}}$,  $\mathrm{\theta_{23}}$ and $\mathrm{\theta_{24}}$ peak around 65$^{\circ}$-75$^{\circ}$ and  75$^{\circ}$-85$^{\circ}$ for 10 eV and 3 eV, respectively. Note that the shifting of  the peak at smaller angles from 65$^{\circ}$-75$^{\circ}$ for 10 eV to 75$^{\circ}$-85$^{\circ}$ for 3 eV shows a tendency towards the {\it pyramid}-consistent angle of 90$^\circ$. Thus, the distributions  in \fig{fig:collisions} (top row) for the $\mathrm{s_{1}}$ ionization route are consistent  with the {\it pyramid}-shape  shown in \fig{fig:break-up} (a). Similarly for the ionization route $\mathrm{s_{2}}$, $\mathrm{C_{s_{2}}(\theta)}$  for $\mathrm{\theta_{23}}$,  $\mathrm{\theta_{24}}$ and $\mathrm{\theta_{34}}$ peak around 115$^{\circ}$ while $\mathrm{C_{s_{2}}(\theta)}$ for $\mathrm{\theta_{12}}$,  $\mathrm{\theta_{13}}$ and $\mathrm{\theta_{14}}$ peak around 65$^{\circ}$-75$^{\circ}$ and    85$^{\circ}$ for 10 eV and 3 eV, respectively (\fig{fig:collisions} bottom row). These distributions are consistent with the {\it pyramid}-shape  shown in \fig{fig:break-up} (b).
Therefore, for the majority (65\%) of QI events the four electrons escape on the vertices of a {\it pyramid}. 
 
 We now provide further evidence  that if the  three electrons escape on a plane at 120$^{\circ}$ from each other then the preferred escape geometry of the other electron is perpendicular to this plane. We do so analytically by expressing the five-body Hamiltonian in hyperspherical coordinates 
 \begin{equation}
 H=\frac{p_{r}^2}{2}+\frac{\bf{\Lambda}^2}{2R^2}+\frac{C(\bf{\Omega})}{R}
\end{equation}
where $\mathrm{\bf{\Omega}=(\alpha_{1},\alpha_{2}, \alpha_{3}, \theta_{1}, \theta_{2}, \theta_{3}, \theta_{4}, \chi_{1},\chi_{2},\chi_{3}, \chi_{4})}$ contains all angular variables describing the positions of the electrons and $\mathrm{\Lambda}$
is a function of $\mathrm{\bf{\Omega}}$ and all conjugate momenta. The total Coulomb interaction $\mathrm{V=C/R}$ acquires in this form simply an angular dependent charge  $\mathrm{C(\bf{\Omega})}$. The hyperspherical coordinates are given by

 \begin{equation}
 \begin{array}{lcl}
R=\sqrt{r_{1}^2+r_{2}^2+r_{3}^2+r_{4}^2}  &&    \chi_{1}=\phi_{3}-\phi_{1} \\
\alpha_{1}=arctan(\frac{r_{1}}{r_{3}}) &&             \chi_{2}=\phi_{4}-\phi_{1} \\
\alpha_{2}=arctan(\frac{\sqrt{r_{1}^2+r_{3}^2}}{r_{4}}) &&\chi_{3}=\phi_{2}-\phi_{1}\\
\alpha_{3}=arctan(\frac{\sqrt{r_{1}^2+r_{3}^2+r_{4}^2}}{r_{2}}) && \chi_{4}=\phi_{1}+\phi_{2}+\phi_{3}+\phi_{4}\\
\end{array}
\end{equation}
where $\mathrm{\phi_{i}}$, $\mathrm{\theta_{i}}$ are the azimuthal and polar angles of the ith electron.  Focusing on  $\mathrm{s_{1}}$, the TTC is  $\mathrm{r_{1}\approx r_{3} \approx r_{4} \neq r_{2}}$. For simplicity we assume $\mathrm{r_{1}\approx r_{3} \approx r_{4}  << r_{2}}$ resulting in  $\mathrm{\alpha_{3}\approx 0}$  (the opposite case would lead to the same result). We then expand $\mathrm{C(\Omega)}$ in powers of $\mathrm{\alpha_{3}}$.
\begin{equation}
C(\Omega) \approx \alpha_{3}^{-1}\sum_{i=1}^{3}c_{n} \alpha_{3}^n.
\end{equation}

 The lowest order term in $\mathrm{\alpha_{3}}$, is the potential term of the four-body Coulomb problem with $\mathrm{Z=4}$, see Eq. 5.  Thus, the problem of finding a stable configuration 
is that of the three-electron problem with the  solution $\mathrm{\alpha_{1}^{*}=\pi/4}$, $\mathrm{\alpha_{2}^{*}=arctan(\sqrt{2})}$, $\mathrm{\chi_{1}^{*}=2\pi/3}$, $\mathrm{\chi_{2}^{*}=4\pi/3}$ and $\mathrm{\theta_{1}=\theta_{3}=\theta_{4}=90^{\circ}}$ \cite{ewr08}. 
These values minimize $\mathrm{c_{2}}$ for any value $\mathrm{\theta_{2}}$. Minimizing $\mathrm{c_{3}}$ with respect to $\theta_{2}$ we find the stable solution $\theta_{2}=0^{\circ}$ which indeed corresponds to 
a {\it pyramid} break-up geometry, which is of lower symmetry than a tetrahedron.

\begin{widetext}
\begin{eqnarray}
c_{0}=&& -\frac{Z}{\sin (\alpha_2)  \sin (\alpha_1)}-\frac{Z}{ \sin (\alpha_2)  \cos (\alpha_1)}-\frac{Z}{ \cos (\alpha_2)} + \frac{1}{  \sin(\alpha_2)\sqrt{1 -\sin(2\alpha_1) \left( \sin\theta_1\sin\theta_3\cos(\chi_1) + \cos\theta_1 \cos\theta_3 \right)}} \nonumber \\
&&+ \frac{1}{\sqrt{ \sin^2 (\alpha_2)  \sin^2 (\alpha_1)+\cos^2 (\alpha_2)-\sin(2\alpha_2)\sin(\alpha_1) \left( \sin\theta_1\sin\theta_4\cos(\chi_2) + \cos\theta_1 \cos\theta_4 \right)}} \nonumber \\
 &&+ \frac{1}{\sqrt{ \sin^2 (\alpha_2)  \cos^2 (\alpha_1)+ \cos^2 (\alpha_2)- \sin(2\alpha_2)\cos (\alpha_1) \left( \sin\theta_3\sin\theta_4\cos(\chi_2-\chi_1) + \cos\theta_3 \cos\theta_4 \right)}} \hspace{2cm} 
\label{eq:5}
\end{eqnarray}
\end{widetext}

 \begin{figure}[h]
  \includegraphics[width=0.49\textwidth]{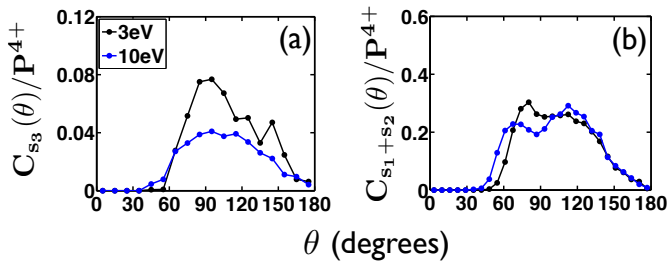}           
 \caption{$\mathrm{C(\theta)}$ for the ionization routes $\mathrm{s_{3}}$ (a) and $\mathrm{s_{1}+s_{2}}$ (b). The lower statistics in (a) compared to (b) dictate using 18 (a)  instead of 28 (b) bins.}
 \label{fig:tetrahedron}
\end{figure}

It is now clear that for the $\mathrm{s_{1}+s_{2}}$ labeled QI events  the four electrons escape on the vertices of a {\it pyramid}, see \fig{fig:collisions}. Since these   events account for roughly 65\% of all QI events the {\it pyramid}-shape prevails for 3 eV and 10 eV excess energy. Why then is the double peak in $\mathrm{C(\theta)}$ more pronounced for 10 eV  than for 3 eV in  \fig{fig:theta}?  One reason is the following:   $\mathrm{C_{s_{1}}(\theta)}$ for $\mathrm{\theta_{12}}$,  $\mathrm{\theta_{23}}$ and $\mathrm{\theta_{24}}$ (\fig{fig:collisions}, top row), and $\mathrm{C_{s_{2}}(\theta)}$ for $\mathrm{\theta_{12}}$,  $\mathrm{\theta_{13}}$ and $\mathrm{\theta_{14}}$ (\fig{fig:collisions}, bottom row) peak at 85$^{\circ}$ for 3 eV while only at 65-75$^\circ$ for 10 eV. In addition,  $\mathrm{C_{s_{1}}(\theta)}$ for $\mathrm{\theta_{13}}$,  $\mathrm{\theta_{14}}$ and $\mathrm{\theta_{34}}$ (\fig{fig:collisions}, top row), and $\mathrm{C_{s_{2}}(\theta)}$ for $\mathrm{\theta_{23}}$,  $\mathrm{\theta_{24}}$ and $\mathrm{\theta_{34}}$ (\fig{fig:collisions}, bottom row) peak roughly at 115$^{\circ}$ for 3 eV and  for 10 eV. Thus, by summing up to obtain $\mathrm{C_{s_{1}+s_{2}}(\theta)}$, see \fig{fig:tetrahedron} (b),  the two peaks are closer for 3 eV (at 85$^{\circ}$ and 115$^{\circ}$) than for 10 eV (at 65$^{\circ}$-75$^{\circ}$ and 115$^{\circ}$) resulting in a stronger overlap and a less pronounced double peak for 3 eV. The same effect is also present  when all QI events are considered in $\mathrm{C(\theta})$ in \fig{fig:theta}.   Another reason  is an ionization route which involves at least four distinct collisions: one collision is $\widehat{12}$ while two of them involve electron 3 and/or  4 each gaining energy by both electrons 1 and 2---we label this route  as  $\mathrm{s_{3}}$. For  $\mathrm{s_{3}}$ TTC is  $\mathrm{r_{1}\approx r_{3} \approx r_{4} \approx r_{2}}$. This spatial distribution is 
 close to the fixed point  of the five-body Coulomb problem corresponding to all four electrons  escaping  on the vertices of a tetrahedron at 109.5$^{\circ}$ from each other. Indeed, in \fig{fig:tetrahedron} (a) we find that $\mathrm{C_{s_{3}}(\theta)}$ has a single peak consistent with a tetrahedron geometry. As expected this single peak becomes sharper with decreasing excess energy; compare  $\mathrm{C_{s_{3}}(\theta)}$ for 10 eV and 3eV in  \fig{fig:tetrahedron} (a).  Thus, when all  ionization routes are considered the contribution of $\mathrm{C_{s_{3}}(\theta)}$ for 10 eV does not smear out the double peak of $\mathrm{C_{s_{1}+s_{2}}(\theta)}$, see \fig{fig:tetrahedron} (b), while it does so for 3 eV.  Further contributing to the difference in the shape of $\mathrm{C(\theta)}$ between 3 eV and 10 eV is that   the \% contribution of  $\mathrm{s_{3}}$  to all QI events increases with decreasing excess energy from 7\% for 10 eV to 11\% for 3 eV.  It is interesting to note that while the tetrahedron does not prevail in the break-up geometry, as generally expected, it is nevertheless present.

In conclusion, we have shown that a {\it pyramid} is the prevailing break-up pattern for QI by single-photon absorption from the ground state of Be for excess energies as low as 3 eV above  threshold.  This pattern can be verified by future quantum mechanical and experimental studies of differential cross sections. Such studies have already been performed for three-electron atoms, see for example \cite{Colgan, ColganA, ren08}.  From our previous results on triple  ionization \cite{ewr08} and our current on QI we conjecture that the four-electron break-up pattern is also initial state dependent. That is, a tetrahedron will be the break-up pattern for initial states where 3 electrons occupy orbitals with similar spatial distribution.  Interestingly, we have shown that the {\it pyramid}-shape is in accord with a classification scheme of {\it momentum transferring attosecond collisions}. This scheme was proven to elucidate correlated multi-electron escape in intra-atomic processes, see also \cite{Emmanouilidou2006JPB, Emmanouilidou2007PRA1}. In doing so it allows for a new perspective on how to time-resolve collisional multi-electron escape dynamics, a problem of intense interest in Attoscience. 



\end{document}